\documentclass[adp,a4paper,fleqn%
]{w-art}
\usepackage{times,cite,w-thm}
\theoremstyle{plain}

\theoremstyle{definition}

\usepackage[]{graphicx}
\usepackage{color}
\usepackage{amssymb,amsmath}
\usepackage{ifthen}
\usepackage{xspace}
\newcommand{\mHt}[2]{{H_1^{#1#2}}\xspace}
\newcommand{\mcHt}[2]{{{\mathcal H}_1^{#1#2}}\xspace}

\newcommand{\bra}[1]{{\langle #1 |}\xspace}
\newcommand{\ket}[1]{{| #1 \rangle}\xspace}
\newcommand{\braket}[2]{{\langle #1| #2 \rangle}\xspace}
\newcommand{\op}[3]{%
\protect{%
\ifthenelse{%
\equal{#2}{}%
}%
{%
\ifthenelse{\equal{#3}{}%
}%
{\ensuremath{{\mathbf #1}}\xspace}%
{\ensuremath{{\mathbf #1}^{#3}}\xspace}}%
{%
\ifthenelse{\equal{#3}{}%
}%
{\ensuremath{{\mathbf #1}_{\mathrm{#2}}}\xs pace}%
{\ensuremath{{\mathbf #1}_{\mathrm{#2}\,#3}}\xspace}}%
}%
}
\newcommand{\adjop}[3]{
\protect{%
\ifthenelse{\equal{#2}{}}%
{\ifthenelse{\equal{#3}{}}%
{\ensuremath{{\mathbf #1}^\dagger}\xspace}%
{\ensuremath{{\mathbf #1}^\dagger_{#3}}\xspace}}%
{\ifthenelse{\equal{#3}{}}%
{\ensuremath{{\mathbf #1}^\dagger_{\mathrm{#2}}}\xspace}%
{\ensuremath{{\mathbf #1}^\dagger_{\mathrm{#2}\,#3}}\xspace}}%
}%
}%

\newcommand{\menge}[1]{\ensuremath{\mathbb{#1}}}
\def\pt{perturbation theory }
\def\BW{Brillouin-Wigner }
\def\RS{Rayleigh-Schr\"odinger }
\def\ecf{E_{\mathnormal CF}}

\newcommand{\tildeOp}[3]{%
\protect{%
\ifthenelse{%
\equal{#2}{}%
}%
{%
\ifthenelse{\equal{#3}{}%
}%
{\ensuremath{{\tilde{\mathbf #1}}}\xspace}%
{\ensuremath{{\tilde{\mathbf #1}}_{#3}}\xspace}}%
{%
\ifthenelse{\equal{#3}{}%
}%
{\ensuremath{{\tilde{\mathbf #1}}_{\mathrm{#2}}}\xspace}%
{\ensuremath{{\tilde{\mathbf #1}}_{\mathrm{#2}\,#3}}\xspace}}%
}%
}

\begin{document}
\DOIsuffix{theDOIsuffix}
\Volume{16}
\Month{01}
\Year{2007}
\pagespan{1}{}
\Receiveddate{XXXX}
\Reviseddate{XXXX}
\Accepteddate{XXXX}
\Dateposted{XXXX}
\keywords{Nearly degenerate perturbation theory, magnetic transitions, strongly correlated electrons.}



\title[Magnetic transitions]{Magnetic transitions in strong coupling
  expansions for nearly degenerate states}
\author[R. Fr\'esard]{Raymond Fr\'esard\inst{1,}%
\footnote{Corresponding author\quad E-mail:~\textsf{Raymond.Fresard@ensicaen.fr},
            Phone: +33-(0)231 45 26 09,
            Fax: +33-(0)231 95 16 00}}
\address[\inst{1}]{Laboratoire CRISMAT, UMR CNRS-ENSICAEN 6508, 6 Boulevard
  Mar\'echal Juin, 14050 Caen Cedex, France}
\author[Ch. Hackenberger]{Christian Hackenberger\inst{2}}
\address[\inst{2}]{Center for Electronic Correlations and Magnetism, Experimentalphysik VI, Institute of Physics, Universit\"at Augsburg, 86135 Augsburg, Germany}
\author[T. Kopp]{Thilo Kopp\inst{2}
}

\begin{abstract}
A strong coupling expansion for a two-band Hubbard model on two sites with
nearly degenerate states is considered. A comparative analysis is performed
for different schemes of perturbation theory which are applicable to systems
with nearly degenerate states. A fourth order approach which builds on a 
four-dimensional low-energy subspace with nearly degenerate states captures
accurately the transition from an antiferromagnetic to a ferromagnetic ground
state at large on-site Coulomb interaction.
\end{abstract}
\maketitle                   


\section{Introduction}
\label{sec:1} 
The emergence and antagonism of ferromagnetic and antiferromagnetic ground states 
in strongly correlated electron systems has always been considered to be
of prime importance for a thorough understanding of the magnetic states
in narrow band systems such as the transition metal oxides.
In a large number of cases the nature of the magnetic coupling is successfully
predicted by appropriate spin-orbital models. Formally the magnetic coupling
results from second order expansion in some hopping amplitude parameter $t$,
and therefore applies to systems in the strong 
coupling regime with strong on-site Coulomb interaction $U$ (with respect to the kinetic
energy $t$, i.e., $t \ll U$). 
In that case spin-orbital models provide a framework to analyze a wealth of
properties (for a recent reference see Ole\'s {\it et al.}~\cite{Oles05} and
references therein).  Nevertheless, many systems are rather tracked down to be
in the intermediate coupling regime, where, for example,  
phase transitions between ferromagnetic and antiferromagnetic phases can take
place (see, e.~g., \cite{Bog10}). 
Such situations are not covered by the standard
spin-orbital models, and higher orders in the expansion in $t$ are
needed. Unfortunately this often implies to take nearly degenerate states into
account. In fact, the determination of the magnetic coupling of nearly degenerate states beyond second order
perturbation theory has been rarely addressed in solid state physics. Our paper aims
at filling this gap.

Whereas small clusters with few electronic states can be exactly diagonalized to characterize 
the electronic state, larger clusters with millions of states require different approaches. 
Already a cluster of three transition metal ions (with six $d$ electrons per site)
and two oxygen ions 
provides more than 600 thousand low energy states which are nearly degenerate. 
To be explicit, this challenge is encountered in the quasi-1D 
Mott-insulator 
Ca$_3$Co$_2$O$_6$, when effective interaction parameters are calculated from a minimal cluster 
of three Co sites (only counting the $3d$ states) and two O ligand sites (exclusively considering $2p$ states),  and a microscopic analysis of the dominant 
ferromagnetic coupling between high-spin next-nearest neighbor Co$^{3+}$ sites has not yet included all
low-energy states. In such a system one would have to implement a perturbation theory 
of at least 5$^{\rm th}$ order (due to the necessity to include ring exchange). Moreover, 
an adequate perturbation theory should be fully qualified to cope with a large number of 
nearly degenerate states in the low energy sector.
Such a situation with several or many nearly degenerate low energy states, which are separated
by a large energy scale $U$ from the high energy sector, is typical for multi-band
transition metal oxides and the question arises if well established perturbation theories are available
with which electronic systems with {\it almost} or {\it nearly degenerate ground states} can be reliably handled~\cite{comment1}. 

Schemes for the implementation of a perturbation theory with {\it degenerate} states in the low
energy sector have been known for a long time and are well presented in textbooks on quantum mechanics and review articles
(e.g., see the Refs.~\cite{Schiff,Messiah,Baym,Fulde,Kato49,Loewdin62,Klein73,Takahashi77}). The extension to {\it almost degenerate} states seems to be straightforward with either of two alternative procedures: (i) one formally assigns a degenerate ``unperturbed'' Hamiltonian to those eigenstates which are nearly degenerate~\cite{Messiah,Fulde}. The perturbation expansion is then set up with respect to this artificially constructed, unperturbed degenerate Hamiltonian with the perturbation being comprised of the terms which lift the degeneracy and the original perturbation terms which connect to the high energy sector. This scheme, which introduces an artificial degeneracy, is in general not physically motivated and may produce inferior convergence as we will demonstrate in Sec.~\ref{sec:3}. (ii) Alternatively, one first diagonalizes the unperturbed Hamiltonian jointly with that part of the perturbation which can be projected onto the low energy sector of the Hilbert space (in which the degenerate and almost degenerate states live). The part of the perturbation which connects to the high energy sector can then be treated in standard perturbation theory~\cite{Baym}. However this latter scheme is valid only up to third order where intermediate states in this scheme are exclusively states from the high energy sector. In fourth order, this scheme breaks down because intermediate states are then in turn allowed to be taken from the low energy sector. We will shortly introduce this procedure in the following section. The standard argument implicates the expectation that the degeneracy or near degeneracy is lifted in second order of perturbation theory. However this is extremely implausible for thousands of nearly degenerate states.

A generalization of scheme (ii) actually exists which extends its validity to all orders of the perturbation theory. This generalization, introduced by Brandow in a review article~\cite{Brandow}, is a Brillouin-Wigner (BW) type of perturbation theory, which requires to calculate the system energies self-consistently~\cite{Brandow-comment}. This self-consistency implies that BW expansions do not scale properly with the particle number in any finite order which can impede their use in solid state physics. 

A considerable number of perturbation techniques for nearly degenerate states were developed for purposes specific to quantum chemistry. 
Here we focus on a perturbation expansion which was developed by Lindgren~\cite{Lindgren}. It is of Rayleigh-Schr\"odinger (RS) type and can handle multi-particle problems with nearly degenerate 
ground states. From the formal structure it is more demanding than a BW expansion but it obeys the linked cluster theorem~\cite{Brueckner} and the correct particle number scaling  without adjustments. 

To our knowledge, neither the BW-scheme of Brandow nor the RS-scheme of Lindgren were 
used extensively in solid state physics in order to solve electronic problems with nearly degenerate states. 
In this context, it is of particular interest if these schemes can be employed
effectively to investigate magnetic states of strongly correlated electron
systems. Especially, can we reliably gain magnetic coupling constants if there
is a competition between ferromagnetic and antiferromagnetic correlations and
only higher orders in perturbation theory can settle the sign of the coupling? 

In the following section (Sec.~\ref{sec:2}), we first introduce the concept of
``nearly degenerate perturbation theory'' and then present the schemes of
Brandow and Lindgren for the generalization of BW and RS perturbation theory,
respectively. As already mentioned, these schemes were little used in solid
state physics and it is worthwhile to review the basics. In order to reassess
their applicability in this context, we employ in Sec.~\ref{sec:3} the
different perturbation expansions to a two-site cluster with two orbitals and
the standard set of local interactions, kinetic transfer of electrons,
amended by a crystal field term acting in the two-orbital subspace.
Such a term arises due to small deviations from the
cubic symmetry, generically observed in the perovskites of transition metal oxides. In the Fock space with two electrons, this system
may display a transition from a singlet to a triplet state for intermediate to large on-site 
Coulomb interaction $U$. This transition to a spin-orbital
entangled state is not expected from conventional reasoning 
as it cannot be identified in standard second-order perturbation theory~\cite{weg}.
We investigate the quality of the different perturbation schemes by comparing the 4$^{\rm th}$ order result to the exact solution. We summarize our analysis in the last section.


\section{Formalism}
\label{sec:2}

We consider the time-independent Hamiltonian problem 
\begin{equation}
H\ket{\Psi_n}=E_n \ket{\Psi_n}
\end{equation}
where $H$ is in the Schr\"odinger representation.
The eigenvalues $E_n$ and the eigenfunctions $\Psi_n$ are labeled by $n$.
We assume that the Hamiltonian $H$  can be  separated into
\begin{equation}
H=H_0 + H_1,
\end{equation}
where  $H_0$ is the unperturbed Hamiltonian which defines an exactly solvable problem
and $H_1$ is the perturbation. Accordingly, we assume that $H_0$ can be diagonalized, 
and consider the energy corrections to
the eigenenergy $E^{(0)}_{n}$ corresponding to the eigenstate $\ket{n}$ of
$H_0$. The eigenenergy with corrections up to an appropriate order $k$ will be denoted as  $E^{(k)}_{n}$ and, 
to infinite order, $E_{n}$ represents the exact eigenenergy of $H$.

\begin{figure}[h!]
\begin{center}
\includegraphics*[width=14.2cm]{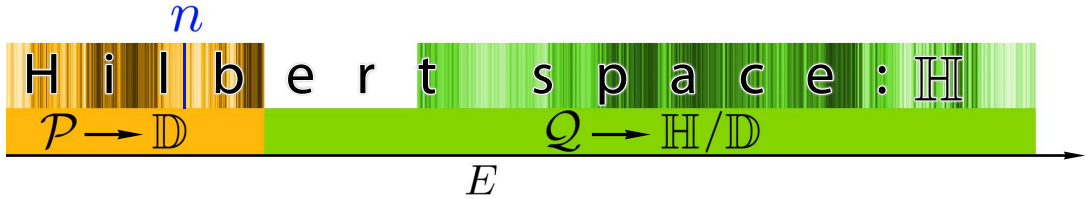}
\end{center}
\caption{(Color online) Structure of the Hilbert space representative for the almost degenerate perturbation theory. It is comprised of a low-energy subspace \menge{D} and a high-energy sector \menge{H/D}. They are separated by an energy gap which is sufficiently large to allow a perturbative expansion in an ``interaction'' which relates states in \menge{D} and \menge{H/D}. The projection operators ${\mathcal P}$ and ${\mathcal Q}$ allow to project general state vectors onto \menge{D} and \menge{H/D}, respectively. The state indexed by $n$ is the reference state for which we consider the perturbation expansion; it is always in \menge{D}.
}
\label{fig:Hilbertspace}
\end{figure}

Moreover, we suppose that the eigenstates of both, the total Hamiltonian $H$
and the unperturbed Hamiltonian $H_0$, can be assigned to either of the two sectors of the
Hilbert space \menge{H}: a low energy sector \menge{D} of dimension $d$ and a high energy sector  
\menge{H/D} which is the orthogonal space of \menge{D} (see Fig.~\ref{fig:Hilbertspace}). 
We note, that the subspace \menge{D} is often referred to
as the ``model space'' (see, e.g., Ref.~\cite{Fulde}). We assume that the states in \menge{D} are all degenerate or
nearly degenerate---with near degeneracy implying that the energy difference of the states in \menge{D} 
is smaller than the difference to any of the states in \menge{H/D}. This decomposition allows
to devise a well-defined perturbation expansion if the perturbative part $H_1$ of the Hamiltonian
has matrix elements smaller than the gap between low energy and high energy states.

The projection operators which are associated with the two complementary subspaces are:
\begin{align}
{\mathcal P} &= \sum_{p \in \menge{D}} \;\;\;\ket{p}\bra{p} \nonumber\\
{\mathcal Q} &= \sum_{q \in \menge{H}/\menge{D}} \ket{q}\bra{q} = 1 -{\mathcal P}
\end{align}

where the index $p$ ($q$) refers to orthonormalized eigenstates $\ket{p}$ ($\ket{q}$) of $H_0$ in the low (high) energy sector of the
Hilbert space. The projections of the eigenvectors of $H$ into the low energy sector are denoted by  $\ket{\Psi_p^{\menge{D}}}={\mathcal P}\ket{\Psi_p}$. 
This notation will be kept for the rest of the paper. We use the normalization
with $\langle \Psi_p \ket{\Psi_p^{\menge{D}}}=1$.

\subsection{Third order perturbation theory for nearly degenerate states}
\label{sec:2.1}

We rewrite the Hamiltonian:
\begin{equation}
\label{lPhys_PT_equation3}
H=H_0+ ({\mathcal P} +{\mathcal Q})H_1({\mathcal P}+{\mathcal Q}) = H_0 + {\mathcal P}\:H_1\:{\mathcal P} +  {\mathcal Q}\:H_1\:{\mathcal P} +  {\mathcal P}\:H_1\:{\mathcal Q} +  {\mathcal Q}\:H_1\:{\mathcal Q}\;.
\end{equation}
At this point we will simply redefine the unperturbed part of the Hamiltonian and the perturbation. With the definitions:
\begin{eqnarray}
\label{lPhys_PT_equation3a}
{\mathcal H}_0&\equiv&H_0{}+{\mathcal P}\,H_1\,{\mathcal P}\;,\\
{\mathcal H}_1&\equiv&{\mathcal P}\,H_1\,{\mathcal Q}+{\mathcal Q}\,H_1\,{\mathcal P}+{\mathcal Q}\,H_1\,{\mathcal Q}\nonumber\;,
\end{eqnarray}
the perturbative problem is restated in an adequate fashion for nearly degenerate cases~\cite{Baym}: 
${\mathcal H}_0$ is to
be diagonalized in the $d$ dimensional subspace \menge{D} and ${\mathcal H}_1$ constitutes the perturbation
of these ``zero order'' states with energies $\mathcal{E}_p^{(0)} \equiv \bra{p} {\mathcal H}_0\ket{p}$. This reformulation of the problem has an irrefutably beneficial effect: all matrix elements of ${\mathcal H}_1$ connecting two states in \menge{D} are zero by construction; therefore only states in the complement $\menge{H}/\menge{D}$ can be intermediate states in the perturbation expansion up to third order. We apply the textbook formulae of  non-degenerate perturbation theory for the new perturbation Hamiltonian ${\mathcal H}_1$ to find for a state $n \in \menge{D}$

\begin{equation}
\label{Eq:third_order}
 \mathcal{E}_{n}^{(3)}- \mathcal{E}_n^{ (0)} = \sum_{q}\frac{\mcHt{n}{q}\mcHt{q}{n}}{\Delta \mathcal{E}_{q}}+
 \sum_{q_1,q_2}\frac{\mcHt{n}{q_1}\mcHt{q_1}{q_2}\mcHt{q_2}{n}}{\Delta \mathcal{E}_{q_1}\Delta \mathcal{E}_{q_2}} \end{equation}
 
where we use the short hand notation $\Delta \mathcal{E}_{q} \equiv \mathcal{E}_n^{ (0)}-\mathcal{E}_q^{(0)}$ and $\mcHt{q_1}{q_2}= \bra{q_1}{\mathcal H}_1 \ket{q_2}$, and $q_1,q_2$ are indices for states in  \menge{H/D}, exclusively. The first order energy correction $\bra{n} {\mathcal H}_1\ket{n}$ vanishes identically by construction because the state $n$ belongs to the subspace \menge{D}. The first order is already encoded in the setup of ${\mathcal H}_0$, and the suppression of the first order in the subsequent perturbation expansion also simplifies the third order term to the form presented in Eq.~(\ref{Eq:third_order}). Beyond third order, intermediate states $p \in \menge{D}$ intrude and spoil the expansion due to the smallness of $\Delta \mathcal{E}_{p}$ in the denominators. 

In order to overcome this difficulty, one has to recast the perturbation expansion into a form that allows to separate the virtual excitations into the high energy states from the low energy processes in such a way that virtual excitations into intermediate low energy states are excluded. Such a reformulation exists for the \BW \pt as well as for the \RS\ perturbation theory. For the first (BW) we refer to Brandow's approach~\cite{Brandow} and for the second (RS) to Lindgren's treatment~\cite{Lindgren}.


\subsection{Brandow's generalized \BW \pt } 
\label{sec:2.2}

The \BW \pt is straightforwardly generalized to account for a low energy subspace \menge{D} of dimension $d$ with nearly degenerate or exactly degenerate states: the states of the model space \menge{D} are excluded from the intermediate summations in the perturbation series. In fact, this procedure is an exact resummation as shown in Sec.~II of Brandow's review~\cite{Brandow} and in Ref.~\cite{Lindgren}. For convenience we label this procedure by BBW (Brandow Brillouin-Wigner) as  Brandow seems to have been the first to make this approach explicit for a ``quasi-degenerate'' model space (see also Refs.~\cite{Bloch,Loewdin62,DesCloizeaux}). 

In the standard BW approach the energy eigenvalues in \menge{D} are found self-consistently from
\begin{equation}
\label{Eq:BBW_energy}
 E_{n} = E_n^{(0)}+\,H_1^{nn} +  \,\sum_{q \in \menge{H/D}}\,\frac{H_1^{nq} H_1^{qn}}{E_{n}- E_q^{(0)}} \,+\,
 \sum_{q_1,q_2 \in \menge{H/D}}\frac{H_1^{nq_1} H_1^{q_1q_2} H_1^{q_2n}}{(E_{n}- E_{q_1}^{(0)})(E_{n}- E_{q_2}^{(0)})} \,
 +\,\cdots \end{equation}
if \menge{D} consists of a single state.
Here, we use the short hand notation  $\mHt{q_1}{q_2}= \bra{q_1}H_1 \ket{q_2}$.

For the BBW scheme with  a $d$-dimensional model space \menge{D}, an effective Hamiltonian~\cite{Lindgren} is introduced for each eigenvalue $E_{n}$
\begin{equation}
\label{H_eff_BBW}
H_{\rm eff}^{n} = {\mathcal P}\,H_0{}\,{\mathcal P}+{\mathcal P}\,W^{n}
 \end{equation}
where the effective interaction ${\mathcal P}\,W^{n}$ in the low energy sector
is found from  
\begin{equation}
\label{Wn}
W^{n} = H_1{}\,{\mathcal P}+ H_1{}\,{\mathcal T}^{n}W^{n}
 \end{equation}
with the resolvent~\cite{Loewdin68}
\begin{equation}
\label{resolvent}
{\mathcal T}^{n} 
= \sum_{q \in \menge{H/D}}\,\frac{\ket{q}\bra{q}}{E_{n} - E_{q}^{(0)}}
 \end{equation}
which is exclusively defined in the high energy sector \menge{H/D}.
Eq.~(\ref{H_eff_BBW}) allows to write the eigenvalue equations
\begin{equation}
\label{H_eff_EVeq}
H_{\rm eff}^{n}\ket{\Psi_n^{\menge{D}}}=  E_{n} \ket{\Psi_n^{\menge{D}}}. \end{equation}
We emphasize that the states $\ket{q}$ are the unperturbed states of the complementary space \menge{H/D} and they 
specify the effective Hamiltonian through the sum in ${\mathcal T}^{n}$ over the high-energy sector. The $E_{n}$ denote the perturbed eigenvalues in the low-energy sector which is spanned by the corresponding eigenvectors $\ket{\Psi_n^{\menge{D}}}$. The perturbed eigenstates $\ket{\Psi_n^{\menge{D}}}$  are identified through
\begin{eqnarray}
\label{BBW_state}
\ket{\Psi_n^{\menge{D}}} &=& (\,\op{1}{}{}+ \,{\mathcal T}^{n}W^{n} )\, \ket{n}\\
&=& (\,\op{1}{}{}+ \,{\mathcal T}^{n}H_1 + \,{\mathcal T}^{n}H_1{\mathcal T}^{n}H_1
       + \,{\mathcal T}^{n}H_1{\mathcal T}^{n}H_1{\mathcal T}^{n}H_1  +\cdots )\, \ket{n}
 \nonumber
 \end{eqnarray}
 Again, we stress that the resolvent ${\mathcal T}^{n}$ and the effective Hamiltonian $H_{\rm eff}^{n}$ depends on the perturbed energies $E_{n}$:
There is an effective Hamiltonian $H_{\rm eff}^{n}$ for each $E_{n}$,
and the eigenvalue equation Eq.~(\ref{H_eff_EVeq}) has to be solved self-consistently for each $E_{n}$ 
with the corresponding $H_{\rm eff}^{n}$.
 
As the states of the model space \menge{D} are excluded from the intermediate
summations in the perturbation series, it is obvious that infinitely many
terms are missing in the BBW summations with respect to the standard BW
approach. However their respective contribution is generated through the
initial diagonalization in the enlarged model space \menge{D}. If \menge{D}
consists of a single state, BBW reduces to \BW \pt and if \menge{D} covers the
entire Hilbert space \menge{H}, the diagonalization already solves the problem
and the perturbation expansion collapses to the zeroth order term. In 
sec.~\ref{sec:3} we will present a model where an appropriately chosen
\menge{D} reduces the full perturbation expansion to the second order term. In
such a case no further higher order processes can be produced which relate
states in the high energy sector \menge{H/D}: due to the restriction to states
in \menge{H/D} on the summations, the second order term already induces the
exact solution. 

Albeit the simplicity of the Brillouin-Wigner and BBW schemes, a formal
disadvantage has to be faced: The linked cluster theorem does not
apply~\cite{Brandow}, which implies an incorrect scaling with the 
number of particles term by term in the perturbation theory. This
inconsistency is only lifted in infinite order.


\subsection{Lindgren's nearly degenerate Rayleigh-Schr\"odinger perturbation theory}
\label{sec:2.3}

Rayleigh-Schr\"odinger perturbation theory cannot be generalized for a problem with almost degenerate states in the model space \menge{D} by excluding straightforwardly the states of \menge{D}  from the intermediate summations in the perturbation series. Such an approach would be correct if the $d$ states of \menge{D} were completely degenerate. However we want to address the case with $d$ almost degenerate states, where the degenerate subspaces of  \menge{D} may  have the dimensions $d_\alpha$ with $1 \leq d_\alpha \leq d$ and $\sum_\alpha d_\alpha =d$.

In this situation it is convenient to introduce the {\it wave operator} $\Omega$~\cite{Moeller,comment2}, which transforms all unperturbed states in \menge{D} into the exact eigenfunctions of $H$, and expand $\Omega$ order by order~\cite{Lindgren}. This yields the set of the desired $d$ eigenenergies and eigenstates~\cite{rem1} to the appropriate order. As the approach traces back to the Rayleigh-Schr\"odinger perturbation theory, it does not suffer from a lack of extensivity, in contrast to \BW perturbation theory.

One introduces the wave operator $\Omega$ through 
\begin{equation}
\ket{\Psi_p} = \Omega \,\ket{\Psi_n^{\menge{D}}}
\end{equation} 
satisfying $\ket{\Psi_p} = \Omega {\mathcal P} \ket{\Psi_p}$
and
\begin{equation}
\Omega {\mathcal Q} = 0 \; .
\end{equation}
Namely, if one projects any of the $d$ low energy exact eigenstates 
$\ket{\Psi^p}$ 
of the full Hamiltonian onto the low energy Hilbert space \menge{D}, the wave operator
takes care of restoring the ``out-projected'' part of the eigenstate, while
$\Omega$ yields a null result when operating on the complementary
subspace.  

To zeroth order the wave operator is simply the projection operator onto \menge{D}:
\begin{equation} 
\Omega^{(0)}= {\mathcal P}
\end{equation}
Once $\Omega^{(i)}$ is obtained to order $i-1$, the effective Hamiltonian
$H_{\rm eff}^{(i)}$ acting on $\menge{D}$ takes the form~\cite{Lindgren}:
\begin{equation} 
\label{Eq:Heff_Lindgren}
H_{\rm eff}^{(i)} = {\mathcal P} H_0 {\mathcal P}+ {\mathcal P}
H_1 \left( \Omega^{(0)}+\Omega^{(1)}+ \Omega^{(2)}+ \ldots +
\Omega^{(i-1)} \right)
\end{equation}
The solution of the eigenvalue problem
\begin{equation} 
\label{Eq:heff}
H_{\rm eff}^{(i)} \ket{\Psi_n^{\menge{D}}} = E_n^{(i)}\ket{\Psi_n^{\menge{D}}}
\end{equation}
yields the exact eigenenergies $E_n^{(i)}$ (to order $i$) in the low energy space, even though it operates on the eigenstates $\ket{n}$ of $H_0$
in \menge{D}. As the wave operator is unique for the entire low-energy Hilbert space, also the effective Hamiltonian is independent of the state $n$, in contrast to that of the Brillouin-Wigner type Hamiltonian in Sec.~\ref{sec:2.2}.

Starting from Schr\"odinger's equation, Lindgren obtained a recursion formula
for $\Omega^{(l)}$:
\begin{equation} 
\left[\Omega^{(l)}, H_0\right] = {\mathcal Q} H_1
\Omega^{(l-1)} - \sum_{m=1}^{l-1} \Omega^{(l-m)}
H_1 \Omega^{(m-1)}
\end{equation}
The lowest orders are then explicitly obtained as:
\begin{align}
\Omega^{(1)}&= \sum_{q \in \menge{H}/\menge{D}} 
\sum_{p \in \menge{D}} \ket{q}\bra{p} \;
\frac{\bra{q} H_1 \ket{p}}{E^{(0)}_p-E^{(0)}_q} \nonumber\\
\Omega^{(2)}&= \sum_{q \in \menge{H}/\menge{D}} 
\sum_{p \in \menge{D}} \ket{q}\bra{p} \;
\frac{\bra{q} H_1 \Omega^{(1)}{}{} - \Omega^{(1)}{}{} H_1 \ket{p}}{E^{(0)}_p-E^{(0)}_q} \nonumber\\
\Omega^{(3)} &= \sum_{q \in \menge{H}/\menge{D}} 
\sum_{p \in \menge{D}} \ket{q}\bra{p}\;
\frac{\bra{q} H_1 \Omega^{(2)}{}{} - \Omega^{(1)}{}{}
H_1\Omega^{(1)}{}{} - \Omega^{(2)}{}{} H_1 \ket{p}}{E^{(0)}_p-E^{(0)}_q}\nonumber\\
\Omega^{(4)} &= \sum_{q \in \menge{H}/\menge{D}} 
\sum_{p \in \menge{D}} \ket{q}\bra{p}\;
\frac{\bra{q} H_1 \Omega^{(3)}{}{} - \Omega^{(1)}{}{}
H_1\Omega^{(2)}{}{} - \Omega^{(2)}{}{} H_1\Omega^{(1)}{}{} -\Omega^{(3)}{}{} H_1\ket{p}}{E^{(0)}_p-E^{(0)}_q}
\label{Eq:omegs}
\end{align}
With the knowledge of $\Omega^{(i)}$, up to the appropriate order $i-1$, one identifies the effective Hamiltonian $H_{\rm eff}^{(i)}$, Eq.~(\ref{Eq:Heff_Lindgren}),
and determines the eigenvalues $E_n^{(i)}$ in the low-energy sector \menge{D} through Eq.~(\ref{Eq:heff}).

Whereas the conventional approach of Sec.~\ref{sec:2.1} is a straightforward Rayleigh-Schr\"odinger perturbation expansion of the eigenenergies to the appropriate order in ${\mathcal H}_1$ (see Eqs.~(\ref{lPhys_PT_equation3a}) and (\ref{Eq:third_order})), the Lindgren approach is an expansion of the effective Hamiltonian  $H_{\rm eff}^{(i)}$ to order $i$ in $H_1$. The diagonalization of $H_{\rm eff}^{(i)}$ then introduces higher orders beyond $i$ into the respective eigenenergies $E_n^{(i)}$.

\section{A two-site model to test perturbation theories} 
\label{sec:3}

It is now compelling to put the theoretical presentation into practical terms. A
comparative analysis of the different perturbational schemes is realized for a
basic model for which an exact solution is known: the two-band Hubbard
Hamiltonian for interacting $e_g$ electrons on a two-site molecule. 
Such a model may be written as follows

\begin{equation}
 H=H_{\rm kin}  +
H_{\rm int} + H_{\rm CF}.
\label{eq:H_deg}
\end{equation}
For $e_g$ electrons there are two orbital flavors: $|x\rangle\sim x^{2}-y^{2}$
and $|z\rangle\sim 3z^{2}-r^{2}$ forming a basis in the orbital space. 
Accordingly, the kinetic energy for hopping in $y$-direction is parameterized by
\begin{equation}
H_{\rm kin}= \sum_{\rho\rho'\sigma}
    t^{\rho\rho'} (c^{\dag}_{1\rho\sigma}c^{}_{2\rho'\sigma} + h.c.),
    \qquad t^{\rho\rho'}=-\frac{t}{4}\left(\begin{array}{cc} 3
        & \sqrt{3} \\ \sqrt{3} &  1
\end{array} \right),
\label{eq:H_kin}
\end{equation}
where $t$ is an effective $(dd\sigma)$ hopping matrix element, $\rho$
is a band index with entries $\rho=x$ or $z$, while $\sigma$ labels the spin states. 
The electron-electron interactions are described by 
the on-site terms, which we write in the following form \cite{Ole83}
\begin{eqnarray}
H_{\rm int} &= &
    U\sum_{i}\bigl( n^{}_{ix\uparrow}n^{}_{ix\downarrow} +
    n^{}_{iz\uparrow}n^{}_{iz\downarrow}\bigr ) 
    +\bigl(U-\frac{5}{2}J_H\bigr)\sum_{i}n^{}_{ix}n^{}_{iz} \nonumber \\ 
  &-&2J_H\sum_{i}\textbf{S}_{ix}\cdot\textbf{S}_{iz} 
   +J_H\sum_{i}\bigl
   (c^{\dag}_{ix\uparrow}c^{\dag}_{ix\downarrow}
    c^{}_{iz\downarrow}c^{}_{iz\uparrow} +
    c^{\dag}_{iz\uparrow}c^{\dag}_{iz\downarrow}
    c^{}_{ix\downarrow}c^{}_{ix\uparrow} \bigr).
\label{eq:H_int}
\end{eqnarray}
Here $U$ and $J_H$ denote the intra-orbital Coulomb and Hund's 
exchange elements, whereas $n_{i\rho}=\sum_{\sigma}n_{i\rho\sigma}$  
is the electron density at site $i$ in the $\rho=x,z$ orbital state. The term
$H_{\rm CF}$ represents the uniform crystal-field splitting 
\begin{equation}
H_{\rm CF} = \frac{1}{2}\ecf \sum_{i\sigma}
(n_{ix\sigma}-n_{iz\sigma})
\label{eq:H_cf}
\end{equation}
resulting from, for example, a uni-axial pressure along the $z$-axis. 

Considerable interest in models with orbital degeneracy has
been fueled by the investigation of the Mott transitions and their understanding
(see, for example, Ref.~\cite{iso}). As concerns 
the lattice version of this  particular model, the surprising richness of its phase diagram is instructive, 
especially in connection with the
CMR manganites (for a review see \cite{Dag}). For instance, several
antiferromagnetic phases and ferromagnetic phases have been studied in
mean-field approximation on the square lattice \cite{Fre05}, or by exact
diagonalization \cite{Hot04} and DMFT studies \cite{Pru10}. The model
is also believed to be a minimal model for the striped nickelates
\cite{Tra95,freeman08}, and striped phases have been
studied in a number of ways (e.~g.\ Refs.~\cite{Zaa94,Mar06}). 

In the following we consider 
\begin{equation}
H_1 = H_{\rm kin}  
\label{eq:def_H1}
\end{equation}
as a perturbation of the local Hamiltonian
\begin{equation}
H_0 = H_{\rm int} + H_{\rm CF}.
\label{eq:def_H0}
\end{equation}

\begin{figure}[tbp]
\centering
\includegraphics*[width=0.8\textwidth]{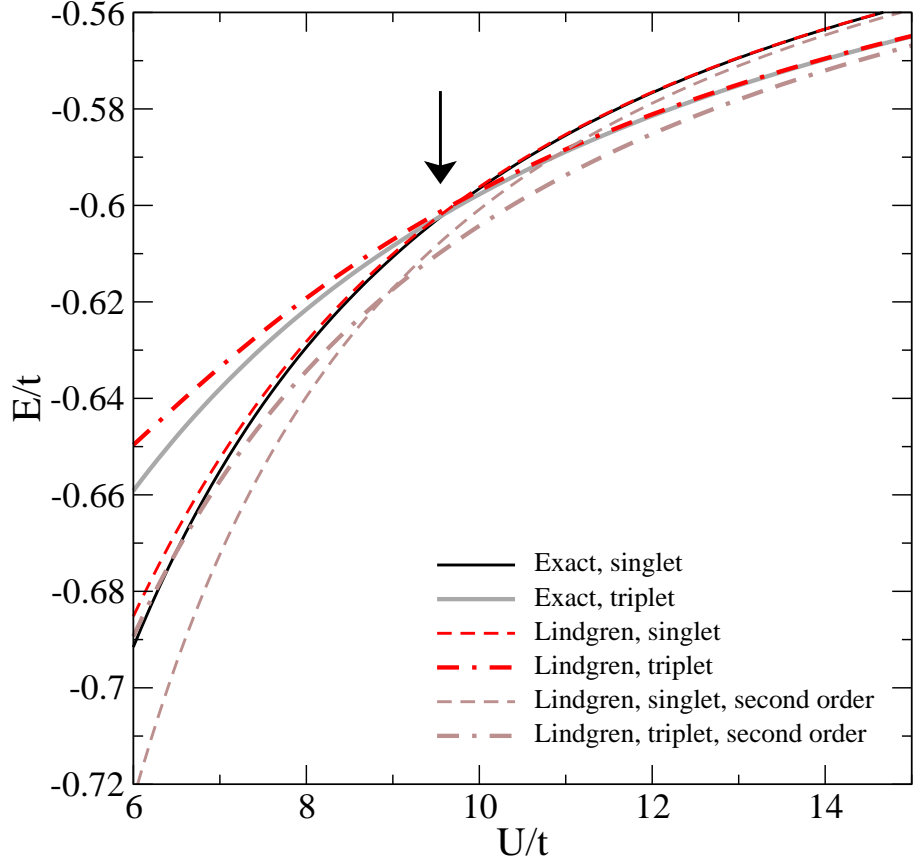} 
\caption{(Color online) Energy of the lowest singlet and triplet
states of the two-band Hubbard model for 2 sites with $J_H/U$ = 0.2, $\ecf/t$ = 0.5. The results for the Lindgren Rayleigh-Schr\"odinger scheme were generated with four nearly degenerate states in the low-energy subspace \menge{D}. The upper two dashed curves display the result to fourth order, the lower two to second order.
}
\label{img:exact_Lindgren}
\end{figure}

\begin{figure}[tbp]
\centering
\includegraphics*[width=0.8\textwidth]{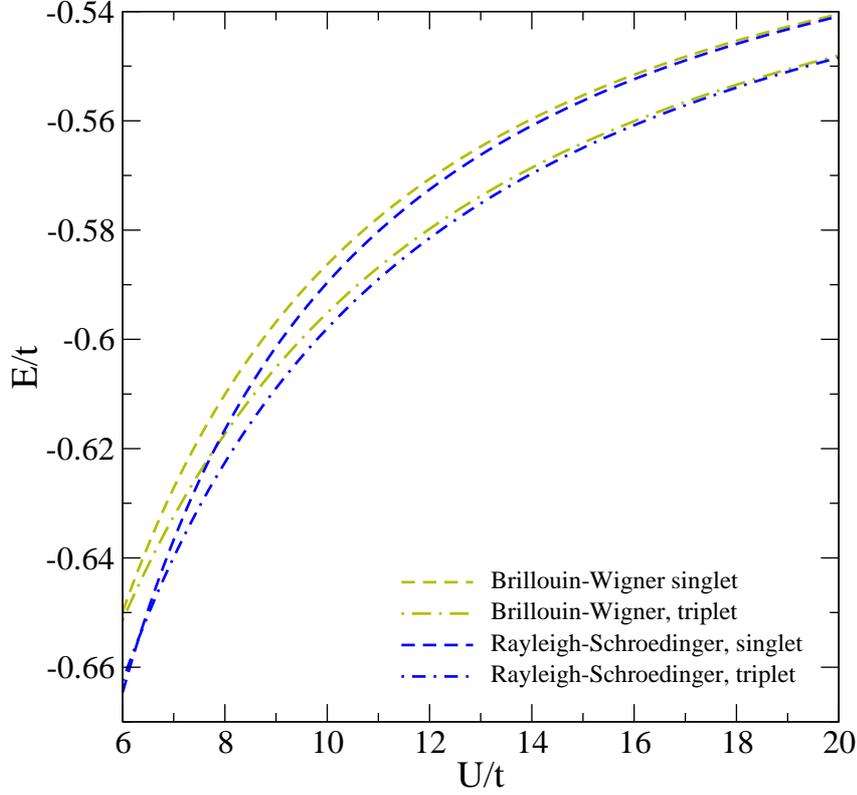} 
\caption{(Color online) Results for the standard Rayleigh-Schr\"odinger and Brillouin-Wigner perturbative expansions for the two-band Hubbard model for 2 sites with $J_H/U$ = 0.2, $\ecf/t$ = 0.5. These evaluations use a one-dimensional low-energy subspace.
}
\label{img:BWandRS}
\end{figure}

With the proper perturbational expressions for the energy corrections,
we can investigate the conditions under which the ground state of the model
(\ref{eq:H_deg}) is ferromagnetic. 
In order to study this transition within the introduced perturbational schemes, one
first needs to determine the set of states forming the subspace
\menge{D}. There are four subspaces, corresponding to states with either
spin 0 (one subspace), or spin 1 (three subspaces). Let us first focus on the triplet
states, restricting ourselves to the ones with $S_z=1$. A basis of
the four-dimensional subspace \menge{D} is given in Appendix A, and it is
easily verified that these basis states are eigenstates of $H_0$. A basis of
the two-dimensional complementary subspace $\menge{H}/\menge{D}$  is also
given in Appendix A. These states are eigenstates of $H_0$ as well. 
We form 
the $2\times 4$ matrix 
$\bra{q} H_1 \ket{p}$. It reads
\begin{equation}\label{eqh29}
\bra{q} H_1 \ket{p} = \frac{t}{4}
\left(\begin{array}{cccc}
0&0&-4&0\\
\sqrt{6}&2&0&-\sqrt{6}
\end{array}\right).
\end{equation}
One can then easily evaluate Eq.~(\ref{Eq:omegs}) and solve Eq.~(\ref{Eq:heff}).
If we now consider the singlet states, the calculation runs in an analogous
fashion. The subspace \menge{D} is again four-dimensional, while the subspace
$\menge{H}/\menge{D}$ is now six-dimensional. Using the basis given in
Appendix B, we obtain the $6 \times 4$ matrix 
$\bra{q} H_1 \ket{p}$ as:
\begin{equation}\label{eqh31}
\bra{q} H_1 \ket{p} =
\frac{t}{4}\left(\begin{array}{cccc}
-\sqrt{2}&-2\sqrt{3}&0&-3\sqrt{2}\\
-\sqrt{2}&0&0&3\sqrt{2}\\
0&0&0&0\\
0&0&2\sqrt{3}&0\\
-\sqrt{6}&-4&0&-\sqrt{6}\\
0&0&2&0
\end{array}\right)
\end{equation}

\begin{figure}[tbp]
\centering
\includegraphics*[width=0.8\textwidth]{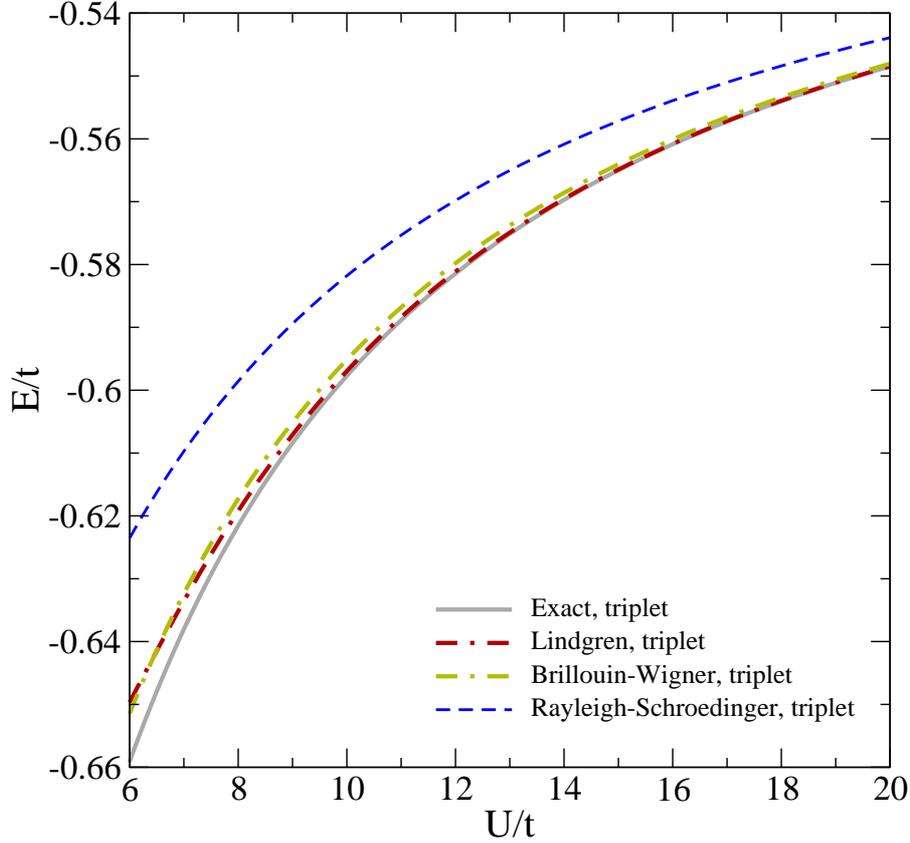} 
\caption{(Color online) Energy of the lowest triplet state of the 2-band Hubbard model for 2~sites with $J_H/U$ = 0.2, $\ecf/t$ = 0.5. The Lindgren approach uses a 4-dimensional low-energy subspace \menge{D}. In contrast, the displayed results for the standard Rayleigh-Schr\"odinger and Brillouin-Wigner perturbative expansions build on a one-dimensional \menge{D}. The three perturbative approaches have been performed up to 4th order.
}
\label{img:Triplet}
\end{figure}

Note that the states $\ket{q}$, which enter Eq.~(\ref{eqh31}), are not eigenstates
of $H_0$, and one further diagonalization step is necessary in order to obtain
the desired  $\bra{q} H_1 \ket{p} =
\sum_{M=1}^{6} \bra{q} H_1 \ket{\varphi^M} \braket{\varphi^M}{p}$.
This is described in Appendix B. Once these steps are worked out, the
desired energy for eigenstate $n$ to perturbative order $i$, viz.\ $E_n^{(i)}\ket{n}$, 
can be evaluated. The higher order terms of the wave operator 
in the Lindgren Rayleigh-Schr\"odinger technique simplify substantially: 
$\Omega^{(2)} = 0 = \Omega^{(4)}$ and
$\Omega^{(3)} =\!\! -\sum_{q \in \menge{H}/\menge{D}} 
\sum_{p \in \menge{D}} \ket{q}\bra{p}\;
{\bra{q}  \Omega^{(1)}{}{} H_1\Omega^{(1)}{}{}  \ket{p}}/({E^{(0)}_p-E^{(0)}_q})$ due to the suppression of high-energy hopping processes.

The result of this evaluation for both singlet and triplet states is shown in
Fig.~\ref{img:exact_Lindgren}. Remarkably, the transition from the
large $U$ ferromagnetic to the intermediate $U$ antiferromagnetic ground state
that occurs at $U_c/t = 9.6$ in the exact diagonalization calculation \cite{weg} (see vertical arrow
in Fig.~\ref{img:exact_Lindgren}) is reproduced in the perturbative treatment, i.e., in Lindgren's Rayleigh-Schr\"odinger type perturbation theory. It occurs here to fourth order perturbation  theory at $U_c^{\rm Lindgren}/t = 9.75$ for four nearly degenerate states in the low energy sector \menge{D}. These states in  \menge{D} are identified by Eqs.~(\ref{S-states}) and (\ref{T-states}) for the singlet and triplet spaces, respectively. For second order perturbation theory with four nearly degenerate states $U_c^{\rm Lindgren}/t = 8.9$ (see Fig.~\ref{img:exact_Lindgren}). The success of the perturbative evaluation derives from the correct identification of the (four-dimensional) low-energy space and the inclusion of the fourth order term, i.e.\ $\Omega^{(3)}$.

\begin{figure}[tbp]
\centering
\includegraphics*[width=0.8\textwidth]{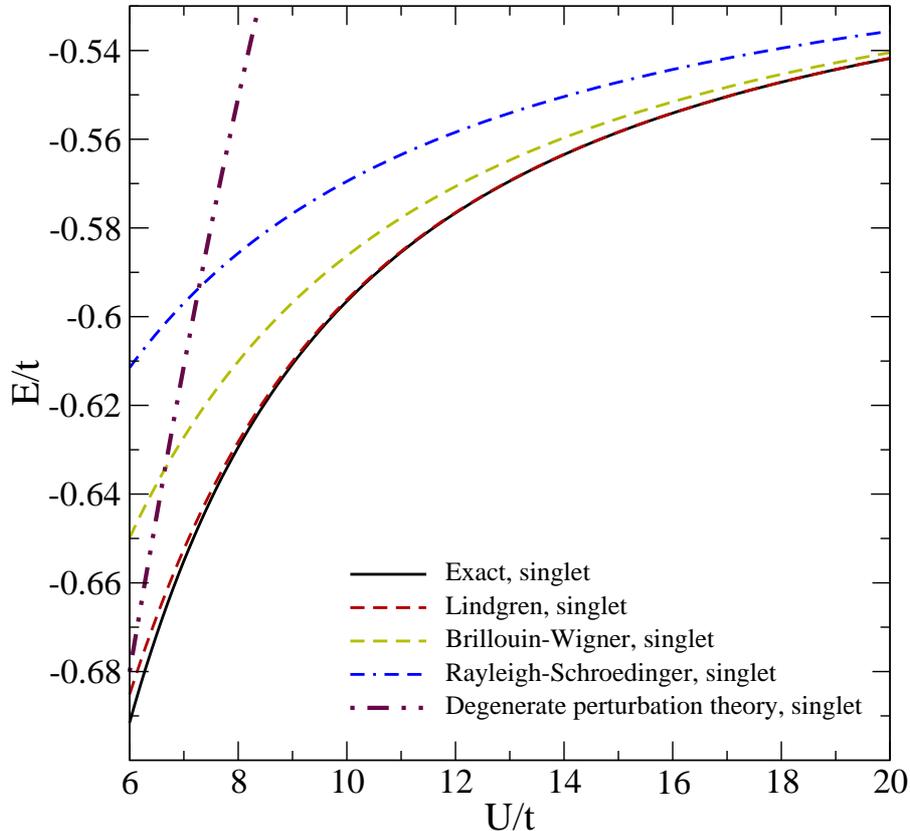} 
\caption{(Color online) Energy of the lowest singlet state of the 2-band Hubbard model for 2~sites with $J_H/U$ = 0.2, $\ecf/t$ = 0.5. Both, the Lindgren approach and the degenerate perturbation theory, take a low-energy subspace \menge{D} of 4 states. In contrast, the displayed results for the standard Rayleigh-Schr\"odinger and Brillouin-Wigner perturbative expansions build on a one-dimensional \menge{D}. All perturbative approaches have been performed up to 4th order.
}
\label{img:Singlet}
\end{figure}

Within non-degenerate Brillouin-Wigner and  Rayleigh-Schr\"odinger
perturbation theory there seems to be no transition for the chosen parameter set (see Fig.~\ref{img:BWandRS} for fourth order perturbation theory). As expected, these results are inferior due to the inappropriately chosen (one-dimensional) subspace \menge{D} (cf.\ to the four-state Lindgren scheme and the exact result in Figs.~\ref{img:Triplet} and \ref{img:Singlet}). An exception is the evaluation of the triplet state within non-degenerate Brillouin-Wigner.

\begin{figure}[tbp]
\centering
\includegraphics*[width=0.8\textwidth]{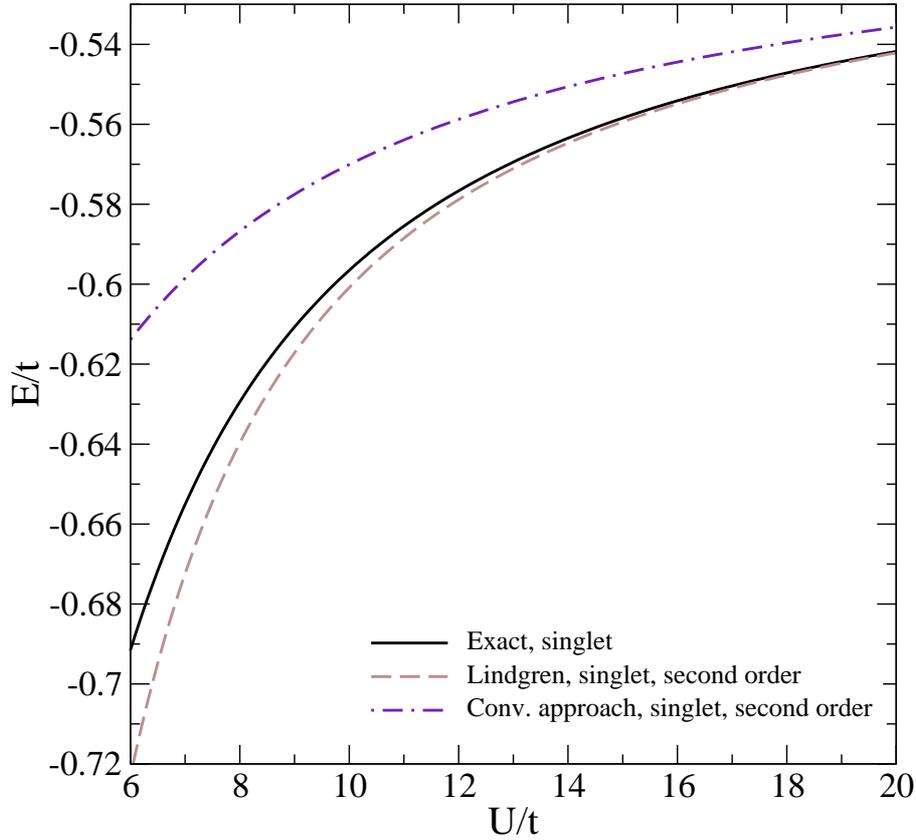} 
\caption{(Color online) Energy of the lowest singlet state of the 2-band Hubbard model for
  2~sites with $J_H/U$ = 0.2, $\ecf/t$ = 0.5. Both, the Lindgren approach and
  the conventional perturbative expansion~\cite{Baym} have been performed up to 2nd order.
}
\label{img:comp_Baym_Lindgren}
\end{figure}

The conventional approach~\cite{Baym}, which was briefly presented in Sec.~\ref{sec:2.1}, reduces to the Rayleigh-Schr\"odinger perturbation theory for this specific model, 
as ${\mathcal H}_0=H_0$ in Eq.~(\ref{lPhys_PT_equation3a}): there are no states in the low-energy sector which may be hybridized through ${\mathcal P}H_1\,{\mathcal P}$. Consequently, the conventional approach compares poorly with the Lindgren approach, even in second order (see. Fig.~\ref{img:comp_Baym_Lindgren}). 

The implementation of degenerate perturbation theory, for example, by setting $\ecf$ to zero and perturbing the four-fold singlet state with respect to $U$ and $\ecf$, which then lifts the degeneracy, has often been suggested. However, the artificial introduction of the zeroth-order degeneracy does not lead to adequate results, as seen in Fig.~\ref{img:Singlet} for the evaluation up to fourth order.

As the matrix elements obey $\bra{q} H_1 \ket{q'} =0$ for the considered model, the Brandow Brillouin-Wigner technique becomes exact for the four nearly degenerate states in the low energy sector \menge{D} (given by  Eqs.~(\ref{S-states}) and (\ref{T-states})) already in second order. This is a special property of the chosen model and not generic for larger systems.

\section{Summary} 
\label{sec:4}

Perturbation theory for atomic clusters is, for typically thousands of low energy states, generically a perturbation expansion with a large number of almost degenerate states. 
Moreover, third or fourth order evaluations are often necessary to attain the required accuracy or include new qualitative effects such as electronic ring exchanges.
Consequently, a perturbation-theory scheme for almost degenerate states is an essential tool in theoretical physics. Almost degenerate perturbation theory has been elaborately investigated in quantum chemistry but, to our knowledge, these approaches have had little impact on solid state modeling. In fact, second or third order theory is well established~\cite{Baym}, but it cannot be straightforwardly extended to fourth order and beyond (see Sec.~\ref{sec:2.1}) due to intruding low-energy states as intermediate states. A much cited technique is to start with the corresponding degenerate problem and then introduce the deviation from the degenerate case as perturbation; such an approach could benefit from the commonly accepted schemes for highly degenerate perturbation theory. However, this approach leads to unsatisfying results in higher order, on account of systematically inaccurate energy denominators---this shortcoming is confirmed for the investigated model (see the result for ``degenerate perturbation theory'' in Fig.~\ref{img:Singlet}). 

A suitable perturbation theory has to start from unperturbed  low- and high-energy states 
which are diagonalized in the respective subspaces. One then has to devise a scheme 
in which the low-energy  subspace is not retraced in the ``string'' of virtual excitation processes induced by the perturbation, otherwise small energy denominators spoil the evaluation. This is achieved in Lindgren's scheme~\cite{Lindgren} for a perturbation theory of Rayleigh-Schr\"odinger type (see Sec.~\ref{sec:2.3}). In this scheme, an effective Hamiltonian is to be identified through appropriate traces over high-energy states, and the eigenvalue problem for the effective Hamiltonian provides the perturbed eigenenergies and states in the low-energy space. For a Brillouin-Wigner type perturbation theory the approach is methodically very simple: the low energy intermediate states are excluded in the perturbation-theory summations, as originally presented by Brandow (see Sec.~\ref{sec:2.2}). The Lindgren scheme can be derived from the Brandow Brillouin-Wigner expressions through expansion~\cite{Brandow}. The schemes were introduced in nuclear physics and quantum chemistry but have not received much attention in solid state physics. 

For the basic example of a two-site two-orbital Hubbard model we presented a comparative analysis of the most prominent perturbative approaches and examined them against a singlet-triplet transition for strong electronic correlations which is known from the exact diagonalization. The Lindgren approach reproduces this transition even quantitatively (in fourth order). The Brandow scheme becomes exact in second order if the subspace of low-energy states \menge{D} is appropriately chosen. However, this property of the Brandow perturbation expansion is particular for the considered model and will generally not persist for larger systems. As to the formal structure, the Lindgren Rayleigh-Schr\"odinger expansion is more demanding but it obeys the linked cluster theorem in each order of the expansion which implies the correct particle-number scaling.

{\it Acknowledgments.}\, R.F. gratefully acknowledges the R\'egion Basse-Normandie and the Minist\`ere de la Recherche for financial support. This work was supported by the Deutsche Forschungsgemeinschaft through TRR 80 (T.K.). C.H. and T.K. are grateful for the warm hospitality at the CRISMAT in Caen where part of this work has been done.

\appendix
\section{Basis of \menge{D} in the triplet subspace}
When considering triplet states (with $S_z=1$), a convenient basis for the
subspace  \menge{D} is given by 
$\{\ket{p}_T,\; 1\leq p \leq 4\}$, with :
\begin{eqnarray}
\ket{1}_T & = & c^{\dagger}_{1,z,\uparrow} c^{\dagger}_{2,z,\uparrow} 
\ket{0} \nonumber \\
\ket{2}_T & = & \frac{1}{\sqrt{2}} \left( c^{\dagger}_{1,x,\uparrow}
    c^{\dagger}_{2,z,\uparrow}  + c^{\dagger}_{1,z,\uparrow}
    c^{\dagger}_{2,x,\uparrow} \right)
\ket{0} \nonumber \\
\ket{3}_T & = & \frac{1}{\sqrt{2}} \left( c^{\dagger}_{1,x,\uparrow}
    c^{\dagger}_{2,z,\uparrow}  - c^{\dagger}_{1,z,\uparrow}
    c^{\dagger}_{2,x,\uparrow} \right)
\ket{0} \nonumber \\
\ket{4}_T & = & c^{\dagger}_{1,x,\uparrow} c^{\dagger}_{2,x,\uparrow}
\ket{0}. 
\label{T-states}
\end{eqnarray}
These states are clearly eigenstates of $H_0$, with eigenvalues $E_1^{(0)} =
-\ecf$, $E_2^{(0)} = E_3^{(0)} = 0$, and $E_4^{(0)} =\ecf$. As for the
complementary space $\menge{H}/\menge{D}$, we use the basis
$\{\ket{q}_T,\; 1\leq q \leq 2\}$  given by:
\begin{eqnarray}
\ket{I}_T & = & \frac{1}{\sqrt{2}} \left( c^{\dagger}_{1,x,\uparrow}
    c^{\dagger}_{1,z,\uparrow}  + c^{\dagger}_{2,x,\uparrow}
    c^{\dagger}_{2,z,\uparrow} \right)
\ket{0} \nonumber \\
\ket{II}_T & = & \frac{1}{\sqrt{2}} \left( c^{\dagger}_{1,x,\uparrow}
    c^{\dagger}_{1,z,\uparrow}  - c^{\dagger}_{2,x,\uparrow}
    c^{\dagger}_{2,z,\uparrow} \right)
\ket{0}.
\end{eqnarray}
Both are eigenstates of $H_0$, with eigenvalues $U-3J_H$.

\section{Basis of \menge{D} in the singlet subspace}
When considering singlet states, a convenient basis for the
subspace  \menge{D} is given by $\{\ket{p}_S,\; 1\leq p \leq 4\}$, with :
\begin{eqnarray}
\ket{1}_S & = & \frac{1}{\sqrt{2}} \left( c^{\dagger}_{1,z,\uparrow}
    c^{\dagger}_{2,z,\downarrow}  - c^{\dagger}_{1,z,\downarrow}
    c^{\dagger}_{2,z,\uparrow} \right)
\ket{0} \nonumber \\
\ket{2}_S & = & \frac{1}{2} \left(c^{\dagger}_{1,x,\uparrow}
    c^{\dagger}_{2,z,\downarrow} - c^{\dagger}_{1,x,\downarrow}
    c^{\dagger}_{2,z,\uparrow} + c^{\dagger}_{1,z,\uparrow}
    c^{\dagger}_{2,x,\downarrow} - c^{\dagger}_{1,z,\downarrow}
    c^{\dagger}_{2,x,\uparrow} 
\right)
\ket{0} \nonumber \\
\ket{3}_S & = & \frac{1}{2} \left(c^{\dagger}_{1,x,\uparrow}
    c^{\dagger}_{2,z,\downarrow} - c^{\dagger}_{1,x,\downarrow}
    c^{\dagger}_{2,z,\uparrow} - c^{\dagger}_{1,z,\uparrow}
    c^{\dagger}_{2,x,\downarrow} + c^{\dagger}_{1,z,\downarrow}
    c^{\dagger}_{2,x,\uparrow} 
\right)
\ket{0} \nonumber \\
\ket{4}_S & = & \frac{1}{\sqrt{2}} \left( c^{\dagger}_{1,x,\uparrow}
    c^{\dagger}_{2,x,\downarrow}  - c^{\dagger}_{1,x,\downarrow}
    c^{\dagger}_{2,x,\uparrow} \right)
\ket{0}. 
\label{S-states}
\end{eqnarray}

These states are eigenstates of $H_0$, with eigenvalues $E_1^{(0)} =
-\ecf$, $E_2^{(0)} = E_3^{(0)} = 0$, and $E_4^{(0)} =\ecf$. As for the
complementary space $\menge{H}/\menge{D}$, we start with the basis
$\{\ket{q}_S,\; 1\leq q \leq 6\}$  given by:
\begin{eqnarray}
\ket{I}_S & = & \frac{1}{2} \left(
  c^{\dagger}_{1,z,\uparrow} c^{\dagger}_{1,z,\downarrow} 
+ c^{\dagger}_{1,x,\uparrow} c^{\dagger}_{1,x,\downarrow} 
+ c^{\dagger}_{2,z,\uparrow} c^{\dagger}_{2,z,\downarrow} 
+ c^{\dagger}_{2,x,\uparrow} c^{\dagger}_{2,x,\downarrow} 
\right)\ket{0} \nonumber \\
\ket{II}_S & = & \frac{1}{2} \left(
  c^{\dagger}_{1,z,\uparrow} c^{\dagger}_{1,z,\downarrow} 
- c^{\dagger}_{1,x,\uparrow} c^{\dagger}_{1,x,\downarrow} 
+ c^{\dagger}_{2,z,\uparrow} c^{\dagger}_{2,z,\downarrow} 
- c^{\dagger}_{2,x,\uparrow} c^{\dagger}_{2,x,\downarrow} 
\right)\ket{0} \nonumber \\
\ket{III}_S & = & \frac{1}{2} \left(
  c^{\dagger}_{1,z,\uparrow} c^{\dagger}_{1,z,\downarrow} 
+ c^{\dagger}_{1,x,\uparrow} c^{\dagger}_{1,x,\downarrow} 
- c^{\dagger}_{2,z,\uparrow} c^{\dagger}_{2,z,\downarrow} 
- c^{\dagger}_{2,x,\uparrow} c^{\dagger}_{2,x,\downarrow} 
\right)\ket{0} \nonumber \\
\ket{IV}_S & = & \frac{1}{2} \left(
  c^{\dagger}_{1,z,\uparrow} c^{\dagger}_{1,z,\downarrow} 
- c^{\dagger}_{1,x,\uparrow} c^{\dagger}_{1,x,\downarrow} 
- c^{\dagger}_{2,z,\uparrow} c^{\dagger}_{2,z,\downarrow} 
+ c^{\dagger}_{2,x,\uparrow} c^{\dagger}_{2,x,\downarrow} 
\right)\ket{0} \nonumber \\
\ket{V}_S & = & \frac{1}{2} \left(
  c^{\dagger}_{1,x,\uparrow} c^{\dagger}_{1,z,\downarrow} 
+ c^{\dagger}_{1,z,\uparrow} c^{\dagger}_{1,x,\downarrow}
+ c^{\dagger}_{2,x,\uparrow} c^{\dagger}_{2,z,\downarrow} 
+ c^{\dagger}_{2,z,\uparrow} c^{\dagger}_{2,x,\downarrow} 
\right)\ket{0} \nonumber \\
\ket{VI}_S & = & \frac{1}{2} \left(
  c^{\dagger}_{1,x,\uparrow} c^{\dagger}_{1,z,\downarrow} 
+ c^{\dagger}_{1,z,\uparrow} c^{\dagger}_{1,x,\downarrow}  
- c^{\dagger}_{2,x,\uparrow}   c^{\dagger}_{2,z,\downarrow} 
- c^{\dagger}_{2,z,\uparrow} c^{\dagger}_{2,x,\downarrow}  
\right)\ket{0}. 
\end{eqnarray}

Unfortunately these eigenstates of $H_{\rm int}$ are not eigenstates
of $H_0$, as ${_S\bra{I}H_0 \ket{II}_S} = {_S\bra{III}H_0 \ket{IV}_S} = -\ecf$,
but the resulting $2\times 2$ blocs are most easily diagonalized. As a result
we obtain the eigenvalues of the unperturbed Hamiltonian as $ E_I^{(0)} =
E_{III}^{(0)} = U - \sqrt{J_H^2 + \ecf^2}$, $ E_{II}^{(0)} =
E_{IV}^{(0)} = U + \sqrt{J_H^2 + \ecf^2}$, and $ E_V^{(0)} =
E_{VI}^{(0)} = U - J_H$.


\begin{thebibliography}{00}
\bibitem{comment1}  In this paper ``almost degenerate'' and ``nearly degenerate'' are to be taken as synonymous expressions.
\bibitem{Brandow-comment} Ref.~\cite{Brandow}, section II; we will not be concerned with the interesting remainder of the article where the energy denominators are expanded to obtain a linked-cluster result.
\bibitem{weg} At first glance, since,
for both vanishing and large crystal field, 
the Kanamori-Goodenough-Anderson
rules predict the  ground state to be antiferromagnetic, ferromagnetism is
unlikely. This, however, is in conflict with recent exact diagonalization
studies that lead to the opposite result for intermediate crystal
field \cite{Bog10}. As shown in Fig.~\ref{img:exact_Lindgren}, the ground state is
ferromagnetic in the strong coupling regime, but antiferromagnetic in the intermediate coupling range.
\bibitem{rem1} As the effective Hamiltonian may be non-hermitian, we here
  refer to the right eigenstates.
\end{thebibliography}
\end{document}